\documentclass[prd,a4paper]{revtex4}
\usepackage{epsfig}

\newcommand{\be}{\begin{equation}}
\newcommand{\ee}{\end{equation}}
\newcommand{\bea}{\begin{eqnarray}}
\newcommand{\eea}{\end{eqnarray}}
\newcommand{\bean}{\begin{eqnarray*}}
\newcommand{\eean}{\end{eqnarray*}}

\newcommand{\gapproxeq}{\lower
.7ex\hbox{$\;\stackrel{\textstyle >}{\sim}\;$}}
\newcommand{\lapproxeq}{\lower
.7ex\hbox{$\;\stackrel{\textstyle <}{\sim}\;$}}

\begin{document}

\bibliographystyle{unsrt}

\title{\bf $\chi_{c0,2}$ decay into light meson pairs and its implication of the scalar meson structures }

\author{Qiang Zhao$^{1,2}$}

\affiliation{1) Institute of High Energy Physics, Chinese Academy
of Sciences, Beijing, 100049, P.R. China}

\affiliation{2)Department of Physics, University of Surrey,
Guildford, GU2 7XH, United Kingdom}

\date{\today}

\begin{abstract}

In light of the recent data from BES collaboration for
$\chi_{c0}\to VV$, $PP$ and $SS$, and from CLEO-c for $\eta\eta$,
$\eta^\prime\eta^\prime$ and $\eta\eta^\prime$, we present a
detailed analysis of the decays of heavy quarkonia into light
meson pairs such as $\chi_{c0,2}\to VV$, $PP$ and $SS$ in a
recently proposed parametrization scheme. An overall agreement
with the data is achieved in $\chi_{c0,2}\to VV$ and $PP$, while
in $\chi_{c0}\to SS$ we find that a possible existence of
glueball-$q\bar{q}$ mixings is correlated with the OZI-rule
violations, which can be further examined at CLEO-c and BESIII in
$\chi_{c0}\to SS$ measurement.

\end{abstract}

\maketitle


\section{Introduction}

The recent systematic measurement of the $\chi_{c0,2}\to VV$, $PP$
and $SS$ by
BES~\cite{bes-99b,bes-04,bes-05,bes-98i,bes-03c,bes-05b} and CLEO
Collaboration~\cite{cleo-c} largely enriches the decay information
about the $\chi_{c0,2}$. A rather unique feature for the light
hadron decay of charmonia is that the transition occurs via
gluon-rich processes. At charmonium mass region, vast
investigations in the literature suggest that non-perturbative QCD
effects are still important and sometimes can become dominant.
Through the study of the hadronic decay of charmonia, one may gain
some insights into the quark-gluon transition mechanisms in the
interplay between non-perturbative  and perturbative QCD. One is
recommended to Ref.~\cite{Brambilla:2004wf} for a detailed review
and prospect of the relevant issues.

Different from the $S$-wave quarkonia, where the annihilation of
the heavy quark and antiquark is a short-distance process, the
pQCD calculation of the $P$-wave quarkonium decays encounters
infrared divergences at order $\alpha_s^3$. For the two photon
decays of $P$-wave charmonia various studies can be found in the
literature~\cite{appelquist-politzer-75,barbieri-76,babcock-rosner-76,barbieri-80,godfrey-isgur-85,bodwin-92,bodwin-95,braaten-lee-03,gupta-96,li-close-barnes-91,lakhina-swanson-06,swanson-review}.
The situation becomes quite complicated in the quarkonium
exclusive hadronic decays, where higher order corrections are no
longer a trivial task~\cite{chernyak}. Attempts were made by
Anselmino and Murgia~\cite{anselmino-94} who found that quark mass
corrections became significant in $\chi_c\to VV$. Some
distinguishable features in the angular distributions of the
final-state-vector-meson decays were also pinned down. More
recently Braguta {\it et al.}~\cite{braguta,luchinsky}
investigated the influence of the internal quark motions on the
scalar and tensor decays into two vectors in the colour-singlet
approximation. Their prediction for $\chi_{c0}\to \omega\omega$
branching ratio was in good agreement with the data, but
significant discrepancies were found for $\chi_{c2}\to
\omega\omega$ compared with the data, which may be due to the
model sensitivity to the choice of the meson structure functions
and possible contributions from the neglected colour-octet
state~\cite{bodwin-92}.

Different roles played by the pQCD transitions and nonperturbative
mechanisms in $\chi_{c0,2}\to \phi\phi$ were studies by Zhou,
Ping, and Zou~\cite{zhou-ping-zou}, who found that the pQCD
calculations for $\chi_{c2}\to \phi\phi$ could reproduce the data,
while the results for $\chi_{c0}\to \phi\phi$ were underestimated.
In contrast, they showed that nonperturbative $^3P_0$ quark pair
creation mechanism could enhance the $\chi_{c0}\to \phi\phi$
branching ratio, but with rather small contributions to
$\chi_{c2}\to \phi\phi$. Their results suggest that
nonperturbative mechanisms are important in $\chi_{c0}\to
\phi\phi$, while pQCD transitions is likely dominant in
$\chi_{c2}\to \phi\phi$.

All these still-controversial observations make the study of the
exclusive decay of $\chi_{c0,2}\to VV$, $PP$, and $SS$ extremely
interesting. Since the decay of $\chi_{c0,2}$ into light hadrons
is via the so-called singly OZI disconnected processes (SOZI), the
study of $\chi_{c0,2}\to VV$, $PP$ and $SS$ will shed light on the
OZI-rule violation phenomena, which are generally driven by
nonperturbative mechanisms. Nonetheless, in the
isoscalar-meson-pair decay channel, the doubly OZI disconnected
process (DOZI) may also contribute. The role played by the DOZI
processes and their correlations with the production mechanisms of
isoscalar scalar meson $f_0$ states are an interesting issue in
the study of the structure of the light scalar mesons at 1$\sim$ 2
GeV, i.e., $f_0(1370)$, $f_0(1500)$, $f_0(1710)$, and $f_0(1810)$.

In this work, we shall present a systematic analysis of the
exclusive decays of $\chi_{c0,2}\to VV$, $PP$ and $SS$ based on an
improved parametrization scheme proposed
recently~\cite{zhao-chic-05}. In light of the new data from
BES~\cite{bes-05b} and CLEO-c Collaboration~\cite{cleo-c}, we
shall identify the role played by the DOZI processes, and gain
some insights into the scalar structures in $\chi_{c0,2}\to SS$.

The content is organized as follows: In Section II, the
parametrization scheme for $\chi_{c0,2}\to MM$ is summarized. In
Section III, we present the analysis and numerical results for
$\chi_{c0,2}\to MM$ in line with the most recent data from BES and
CLEO-c. A short summary will be given in Section IV.

\section{Parametrization for $\chi_{c0,2}\to MM$}

In Ref.~\cite{zhao-chic-05} the decay of $\chi_{c0,2}\to VV$, $PP$
and $SS$ was investigated in a parametrization scheme where the
production of the final state hadrons were described by a set of
transition amplitudes for either SOZI or DOZI processes. Such a
parametrization as a leading order approximation is useful for
identifying the roles played by different transition mechanisms
and will avoid difficulties arising from our poor knowledge about
the nonperturbative dynamics. Associated with the up-to-date
experimental data, we can constrain the model parameters and make
predictions which can be tested in future measurements.

The detailed definition of the parametrization was given in
Ref.~\cite{zhao-chic-05}, we only summarize the main ingredients
here with slightly rephrased expressions:

i) The basic transition amplitude is defined to be the $c\bar{c}$
annihilation into two gluons which then couple to two non-strange
quark pairs to form final state mesons:
\be
\label{sozi}\langle (q_1\bar{q_2})_{M1} (q_3\bar{q_4})_{M2} | V_0
| \chi_c\rangle \equiv g_{\langle 14\rangle}g_{\langle
23\rangle}\equiv g_0^2 \ ,
\ee
where $V_0$ is the interaction potential, and $q(\bar{q})$ is
non-strange quark (antiquark) with $g_{\langle
14\rangle}=g_{\langle 23\rangle}= g_0$. Basically, such a coupling
will depend on the quantum numbers of the initial quarkonium. We
separate the partial decay information by introducing a
conventional form factor in the calculation, i.e., ${\cal F}(|{\bf
p}|)\equiv |{\bf p}|^{2l}\exp(-|{\bf p}|^2/8\beta^2)$ with
$\beta=0.5$ GeV,  for the relative $l$-wave two-body decay.

ii) To include the SU(3) flavour symmetry breaking effects, we
introduce
\be
R \equiv  \langle (q\bar{s})_{M1}(s\bar{q})_{M2}|V_0|\chi_c\rangle
/ g_0^2 = \langle (s\bar{q})_{M1}(q\bar{s})_{M2}|V_0|\chi_c
\rangle / g_0^2,
\ee
which implies the occurrence of the SU(3) flavour symmetry
breaking at each vertex where a pair of $s\bar{s}$ is produced,
and $R=1$ is in the SU(3) flavour symmetry limit. For the
production of two $s\bar{s}$ pairs via the SOZI potential, the
recognition of the SU(3) flavor symmetry breaking in the
transition is accordingly
\be
R^2 = \langle (s\bar{s})_{M1}(s\bar{s})_{M2}|V_0|\chi_c\rangle /
g_0^2 \ .
\ee

iii) The DOZI process is parametrized by introducing parameter $r$
accounting for its relative strength to the SOZI amplitude:
\be
r\equiv \langle (s\bar{s})_{M1}(q\bar{q})_{M2}|V_1|\chi_c\rangle /
g_0^2 = \langle (q\bar{q})_{M1}(s\bar{s})_{M2}|V_1|\chi_c\rangle /
g_0^2,
\ee
where $V_1$ denotes the interaction potential.

iv) Scalar glueball  state can be produced in company with an
isoscalar $q\bar{q}$ or in pair in the final state. We parametrize
their amplitudes by introducing an additional quantity $t$ for the
relative strength of the process of glueball production recoiling
a $q\bar{q}$ to the basic amplitude $g_0^2$:
\bea
 \langle (q\bar{q}) G| V_2| \chi_c\rangle & \equiv & t \langle
(q\bar{q})_{M1} (q\bar{q})_{M2} | V_0 | \chi_c\rangle  = t g_0^2 \
.
\eea
A reasonable assumption for the glueball coupling is that the
glueball does not pay a price to couple to $gg$, namely, the
so-called ``flavor-blind assumption" following the gluon counting
rule. Under such a condition, parameter $t$ has a value of unity,
and the glueball production amplitude is of the same strength as
the basic amplitude $g_0^2$. Similarly, the production of a
glueball pair can be expressed as
\be
\langle GG| V_3| \chi_c\rangle  =  t  \langle (q\bar{q}) G| V_2|
\chi_c\rangle = t^2 g_0^2 \ .
\ee

Considering a general expression for isoscalar meson pair
production with $q\bar{q}$ and glueball components, e.g. $M_{1,2}=
x_{1,2}|G\rangle + y_{1,2} |s\bar{s}\rangle +z_{1,2}
|n\bar{n}\rangle$, we can write the transition amplitude for
$\chi_c\to M_1 M_2$ as
\bea
&&\langle M_1(I=0) M_2(I=0) | (V_0+ V_1 +V_2 +V_3)|\chi_c\rangle
\nonumber\\
&=& \langle (x_1 G + y_1 s\bar{s} + z_1 n\bar{n})(x_2 G + y_2
s\bar{s} + z_2 n\bar{n})|(V_0+ V_1 +V_2 +V_3)|\chi_c\rangle \nonumber\\
&= & g_0^2 [(x_1 t( t x_2 + R y_2 + \sqrt{2} z_2) + y_1 R (
t x_2 + (1+r) R y_2 + \sqrt{2} r z_2) \nonumber\\
&&+ z_1(\sqrt{2} t x_2 +\sqrt{2} r R y_2 +(1+2r) z_2)] \ .
\eea
For meson pair production with isospin $I=1/2$ and 1, the
transitions only occur via potential $V_0$, and they can be
expressed as
\bea
\langle M_1(I=1/2) M_2(I=1/2)| V_0 |\chi_c\rangle & = & R g_0^2 \
,\\
\langle M_1(I=1) M_2(I=1)| V_0 |\chi_c\rangle & = & g_0^2 \ .
\eea

The modification of the above parametrization rule compared to
Ref.~\cite{zhao-chic-05} is on the glueball production. Here,
parameters $r$ and $t$ are explicitly separated out.
 Parameter $r$ describes the property of the $q\bar{q}$-$gg$
couplings in the DOZI processes. Apparent contributions from the
DOZI processes generally demonstrate the importance of the
OZI-rule violations due to long-range
interactions~\cite{geiger-isgur}. In contrast, parameter $t$
distinguishes the $G$-$gg$ coupling from the $q\bar{q}$-$gg$, and
will allow us to investigate the role played by glueball
productions. In the present scheme the underlying physics denoted
by the parameters can be more clearly identified.

\section{Decay of $\chi_{c0,2}\to MM$}

In this Section we revisit $\chi_{c0,2}\to VV$, $PP$ and $SS$
taking into account the new data from both BES and CLEO-c.

\subsection{$\chi_{c0,2}\to VV$}

For $\chi_{c0,2}\to VV$, three channels, i.e. $\phi\phi$,
$\omega\omega$ and $K^{*0}\bar{K^{*0}}$, have been measured by BES
collaboration~\cite{bes-99b,bes-04,bes-05}. Since we neglect
glueball component in $\omega$ and $\phi$, and assume that
$\omega$ is pure $n\bar{n}$ and $\phi$ is pure $s\bar{s}$ due to
ideal mixing, we can determine parameters $g_0$, $r$, and $R$.
Predictions for $\chi_{c0,2}\to \rho\rho$ and $\omega\phi$ can
then be made.

In Table~\ref{tab-1}, the parameters are presented. In
Table~\ref{tab-2}, we list the fitting results for $\chi_{c0,2}\to
VV$ in comparison with the experimental
data~\cite{bes-99b,bes-04,bes-05}. Also, the result by fitting the
PDG average values for $\chi_{c0,2}\to \phi\phi$, $\omega\omega$
and $K^{*0}\bar{K^{*0}}$ are included.

One apparent feature is that the OZI-rule violation and SU(3)
flavor symmetry breaking are much obvious in $\chi_{c0}\to VV$
than in $\chi_{c2}\to VV$. Parameter $r$ is found to be about
$20\%$ for $\chi_{c0}$, while its central values are about $1\%$
for $\chi_{c2}$ though the uncertainties are about $10\%$. The
consequence of small DOZI process contributions is that the
production branching ratios for $\chi_{c0,2}\to \omega\phi$ become
rather small. For instance, predictions for the branching ratio of
$\chi_{c0}\to \omega\phi$ are at least one order of magnitude
smaller than $\phi\phi$ channel, and the PDG averaged values for
the experimental data lead to a negligibly small branching ratio
for $\chi_{c2}\to \omega\phi$. Further experimental measurement
confirmation of this prediction will be extremely interesting.

The $\rho\rho$ branching ratio turns to be sensitive to the
experimental uncertainties carried by those available data.
Different from other decay channels, which are determined by
parameters $r$, $R$ and $g_0$ in a correlated way, it only depends
on parameter $g_0$. Therefore, the $\rho\rho$ channel is ideal for
testing this parametrization scheme, and can put further
constraint on the parameters.


\subsection{$\chi_{c0,2}\to PP$}

Decay channels of $\chi_{c0,2}\to \eta\eta$, $K^+K^-$,
$K_s^0K_s^0$ and $\pi\pi$ have been measured at
BES~\cite{bes-99b,bes-98i,bes-03c,bes-05b}. However, as studied in
Ref.~\cite{zhao-chic-05}, the relatively large uncertainties with
$\chi_{c0}\to \eta\eta$ brought significant errors to parameter
$r$, and the role played by the DOZI processes cannot be
clarified. It was shown in Ref.~\cite{zhao-chic-05} that within
the uncertainties of $BR_{\chi_{c0}\to\eta\eta}=(2.1\pm 1.1)\times
10^{-3}$~\cite{bes-05b}, the relative branching ratios of
$\chi_{c0,2}\to\eta\eta$, $\eta\eta^\prime$ and
$\eta^\prime\eta^\prime$ were very sensitive to the OZI-rule
violation effects, and the branching ratio fractions can vary
drastically. The world averaged data for $\chi_{c0}\to K^+ K^-$,
$K_s^0K_s^0$, and $\pi\pi$~\cite{pdg2006} do not deviated
significantly from the BES
data~\cite{bes-99b,bes-98i,bes-03c,bes-05b} except that
$BR_{\chi_{c0}\to\eta\eta}=(1.9\pm 0.5)\times 10^{-3}$ has much
smaller errors.
 Recently, CLEO-c publishes their results for
$\chi_{c0,2}\to \eta\eta$, $\eta^\prime\eta^\prime$ and
$\eta\eta^\prime$~\cite{cleo-c}, with
$BR_{\chi_{c0}\to\eta\eta}=(3.1\pm 0.5\pm 0.4\pm 0.2)\times
10^{-3}$, $BR_{\chi_{c0}\to\eta^\prime\eta^\prime}=(1.7\pm 0.4\pm
0.2\pm 0.1 )\times 10^{-3}$ and
$BR_{\chi_{c0}\to\eta\eta^\prime}<0.5\times 10^{-3}$. Upper limits
are given for $\chi_{c2}$, i.e. $BR_{\chi_{c2}\to\eta\eta}<
0.47\times 10^{-3}$, $BR_{\chi_{c2}\to\eta^\prime\eta^\prime}<
0.31\times 10^{-3}$, and $BR_{\chi_{c2}\to\eta\eta^\prime}<
0.23\times 10^{-3}$.

Adopting the world-average data from PDG~\cite{pdg2006} and
including the new data from CLEO-c~\cite{cleo-c}, we can now make
a constraint on the model parameters for $\chi_{c0}\to PP$. We
also make a fit for $\chi_{c2}\to PP$ in a similar way with the
experimental bound limits. The fitted parameters and branching
ratios are listed in Table~\ref{tab-3} and \ref{tab-4},
respectively.

It shows that the decay of $\chi_{c0}\to PP$ can be described
consistently with small $\chi^2$. A prominent feature is that the
SU(3) flavor symmetry breaking effects turn out to be small, i.e.
$R=1.035\pm 0.067$ does not deviate significantly from unity.
Meanwhile, parameter $r=-0.120\pm 0.044$ suggests that
contributions from the DOZI processes are not important. The
production of $\eta\eta^\prime$ is thus strongly suppressed which
is consistent with CLEO-c results~\cite{cleo-c}. These features
indicate that pQCD transitions play a dominant role in $PP$ decay
channels.

In $\chi_{c2}\to PP$, by fitting the PDG data and adopting the
CLEO-c bound limits for $\eta\eta$, $\eta^\prime\eta^\prime$ and
$\eta\eta^\prime$, we obtain results with large $\chi^2$. Contrary
to $\chi_{c0}\to PP$, the fitted parameter $R=0.778\pm 0.067$
indicates significant SU(3) flavor symmetry breakings. The
OZI-rule violation parameter $r=-0.216\pm 0.102$ also suggests
that the DOZI processes are relatively more influential than in
$\chi_{c0}$. However, this could be due to the poor status of the
data. Notice that $BR_{\chi_{c2}\to K^+K^-}=(0.77\pm 0.14)\times
10^{-3}$ and $BR_{\chi_{c2}\to K_s^0K_s^0}=(0.67\pm 0.11)\times
10^{-3}$ have violated the isospin relation drastically. It needs
further experiment to check whether this is due to datum
inconsistency  or unknown mechanisms.

It is interesting to see the change of the branching ratio average
for $K^+K^-$ in the past editions of PDG from 1998 - 2006. PDG1998
quoted $BR_{\chi_{c2}\to K^+K^-}=(1.5\pm 1.1)\times
10^{-3}$~\cite{pdg1998} which was measured by DASP
Collaboration~\cite{dasp-79}. In PDG2000~\cite{pdg2000}, it was
averaged to be $BR_{\chi_{c2}\to K^+K^-}=(0.81\pm 0.19)\times
10^{-3}$ with the measurement from BES Collaboration, $(0.79\pm
0.14\pm 0.13)\times 10^{-3}$~\cite{bes-98i}. In
PDG2004~\cite{pdg2004}, this branching ratio was revised to be
$BR_{\chi_{c2}\to K^+K^-}=(0.94\pm 0.17\pm 0.13)\times 10^{-3}$ by
using $BR(\psi(2S)\to \gamma\chi_{c2})=(6.4\pm 0.6)\%$ and
$BR(\psi(2S)\to J/\psi(1S)\pi^+\pi^-)=0.317\pm 0.011$. Then, in
PDG2006~\cite{pdg2006}, this quantity was revised again to be
 $BR_{\chi_{c2}\to K^+K^-}=(0.77\pm 0.14)\times
10^{-3}$, but without explicit explanations. In contrast to this
is that the branching ratio for $K_s^0K_s^0$ has not experienced
drastic changes. Further experimental investigation of these two
channels will be necessary for understanding the $\chi_{c2}\to PP$
decays.

\subsection{$\chi_{c0,2}\to SS$}

The scalar pair production $\chi_{c0}\to SS \to \pi^+\pi^- K^+\
K^-$ is analyzed at BES~\cite{bes-05b}. The intermediate
$K_0^*\bar{K_0^*}$ pair has a branching ratio of
$(1.05\begin{array}{c}+0.39\\ -0.30\end{array})\times 10^{-3}$ in
its decay into $\pi^+\pi^-K^+ K^-$ and a set of $f_0^i f_0^j$
pairs are measured, where $i, \ j=1, \ 2, \ 3$ denotes
$f_0(1710)$, $f_0(1500)$ and $f_0(1370)$, respectively. The
interesting feature is that the $f_0(1370)f_0(1710)$ pair
production is found to have the largest branching ratio in
comparison with other $f_0$ pairs. Theoretical interpretation for
such an observation is needed and in Ref.~\cite{zhao-chic-05}, a
parametrization for the SOZI and DOZI processes suggests that
glueball-$q\bar{q}$ mixings can lead to an enhanced
$f_0(1370)f_0(1710)$ branching ratio in $\chi_{c0}$ decays.
However, due to the unavailability of the data for other scalar
meson pair decays, estimate of the absolute branching ratios were
not possible. Here, incorporated by the data for
$K_0^*(1430)\bar{K_0^*}(1430)$, we expect to have more
quantitative estimates of the $\chi_{c0,2}\to SS$ branching
ratios.

To proceed, several issues have to be addressed:

i) The scalars, $f_0(1370)$, $f_0(1500)$ and $f_0(1710)$, are
assumed to be mixing states between scalar $q\bar{q}$ and glueball
$G$. On the flavor singlet basis, the state mixing can be
expressed as
\be
\left(
\begin{array}{c}
|f_0(1710)\rangle \\
|f_0(1500)\rangle \\
|f_0(1370)\rangle
\end{array}
\right) =U \left(
\begin{array}{c}
|G\rangle \\
|{s\bar{s}}\rangle \\
|{n\bar{n}}\rangle
\end{array}
\right)= \left(
\begin{array}{ccc}
x_1 & y_1 & z_1 \\
x_2 & y_2 & z_2 \\
x_3 & y_3 & z_3
\end{array}
\right) \left(
\begin{array}{c}
|G\rangle \\
|{s\bar{s}}\rangle \\
|{n\bar{n}}\rangle
\end{array}
\right) \ ,
\ee
where $x_i$, $y_i$ and $z_i$ are the mixing matrix elements
determined by the perturbation
transitions~\cite{close-amsler,close-kirk,close-zhao}. We adopt
the mixing matrix $U$ from Ref.~\cite{close-zhao}:
\bea
\label{mix-2} U= \left(
\begin{array}{ccc}
0.36 & 0.93 & 0.09 \\
-0.84 & 0.35 & -0.41 \\
0.40 & -0.07 & -0.91
\end{array}
\right)  \ .
\eea
In order to examine the sensitivities of the branching ratios to
the scalar meson structures in the numerical calculations, we will
also apply several other mixing schemes~\cite{giacosa,cheng,he-06}
which are different from Ref.~\cite{close-zhao}.

ii) In $\chi_{c0,2}\to VV$ and $PP$ the SU(3) flavor symmetry
breaking turns to be at a magnitude of 10$\sim 20\%$. Namely, the
deviation of the SU(3) flavor symmetry parameter $R$ from unity is
small. Due to lack of data we assume that a similar order of
magnitude of the SU(3) flavor symmetry breaking appears in
$\chi_{c0}\to SS$, and it is natural to assume $R=1$ as a leading
order estimate.

We can thus determine the basic transition strength $g_0$ via
\be
\Gamma(\chi_{c0}\to K_0^*\bar{K_0^*})=\frac{|{\bf p}|g_0^4
R^2{\cal F}(|{\bf p}|)}{4\pi M^2_{\chi_{c0}}} \ ,
\ee
where ${\bf p}$ is the three-vector momentum of the final state
$K_0^*$ in the $\chi_{c0}$-rest frame, and ${\cal F}(|{\bf p}|)$
is the form factor for the relative $l$-wave two-body decay. The
partial decay width $\Gamma(\chi_{c0}\to K_0^*\bar{K_0^*})$ has
been measured by BES~\cite{bes-05b}:
\be\label{kk}
BR(\chi_{c0}\to K_0^*\bar{K_0^*}\to\pi^+\pi^-K^+K^-)=(10.44\pm
1.57\begin{array}{c}+3.05\\ -1.90\end{array})\times 10^{-4} \ ,
\ee
with $BR(K_0^*\to K^+\pi^-) = BR(\bar{K_0^*}\to
K^-\pi^+)=0.465$~\cite{pdg2006}.

iii) Since there is no constraint on the parameter $t$, we apply
the flavor-blind assumption, $t=1$, as a leading order
approximation.

iv) In order to accommodate the BES data~\cite{bes-05b}, we adopt
the same branching ratios for $f_0\to PP$ as used in
Ref.~\cite{close-zhao}:
\bea
BR(f_0(1710)\to \pi\pi)& =& 0.11\times BR(f_0(1710)\to K\bar{K}) =0.11\times 0.6 \ , \\
BR(f_0(1500)\to \pi\pi) &=& 0.349 \ , \\
BR(f_0(1500)\to K\bar{K}) &=& 0.086 \ , \\
BR(f_0(1370)\to K\bar{K}) &=& 0.1\times BR(f_0(1370)\to \pi\pi) =
0.1\times 0.2 \ .
\eea
It should be noted that the final predictions for $\chi_{c0}\to
f_0^i f_0^j\to \pi^+\pi^-K^+K^-$ are sensitive to the above
branching ratios. For the charged decay channel, factor $1/2$ and
$2/3$ will be included in the branching ratio of $f_0\to K^+K^-$
and $\pi^+\pi^-$, respectively. Detailed analysis of the $f_0$
states can be found in Ref.~\cite{bugg-review} and references
therein.

Now, we are left with only one undetermined parameter $r$. By
taking the measured branching ratio~\cite{bes-05b}:
\be\label{ss13}
BR(\chi_{c0}\to f_0(1370)f_0(1710)) \cdot
BR(f_0(1370)\to\pi^+\pi^-)\cdot BR(f_0(1710)\to K^+K^-) = (7.12\pm
1.46\begin{array}{c} +3.28 \\ -1.68\end{array})\times 10^{-4} \ ,
\ee
we determine $r= 1.31\pm 0.19$.  Consequently, predictions for
other $SS$ decay channels can be made and the results are listed
in Table~\ref{tab-5}.

A remarkable feature arising from the prediction is that
$BR(\chi_{c0}\to f_0(1370)f_0(1710))$ turns out to be the largest
one in all the $f_0$ pair productions with the constraint from
$K_0^*(1430)\bar{K_0^*}(1430)$. As listed in Table~\ref{tab-5}
branching ratios of $f_0(1370)f_0(1370)$ and $f_0(1370)f_0(1500)$
are at order of $1\%$. Their signals in $\pi^+\pi^-K^+K^-$ are
suppressed due to their small branching ratios to $\pi^+\pi^-$ and
$K^+K^-$~\cite{wa102,bes-phi,bes-plb}. As a comparison decay
channels with $f_0(1710)\to K^+K^-$ are less suppressed. Apart
from the dominant channel $f_0(1370)f_0(1710)$, our calculation
shows that $\chi_{c0}$ has also large branching ratios into
$\pi^+\pi^-K^+K^-$ via $f_0(1500)f_0(1710)$. It shows that our
results for $\chi_{c0}\to f_0^if_0^j\to\pi^+\pi^-K^+K^-$ provide a
consistent interpretation for the BES data~\cite{bes-05b} though
some of the predictions strongly depend on the estimates of the
branching ratios of $f_0\to \pi^+\pi^-$ and $K^+K^-$.

The value of $r=1.31\pm 0.19$ suggests an important contribution
from the DOZI processes in $\chi_{c0}\to f_0^if_0^j$, which is
very different from the results in $VV$ and $PP$ channels. This
certainly depends on the mixing matrix for the scalars, and also
correlated with parameters $R$ and $t$. At this moment, we still
lack sufficient experimental information to constrain these
parameters simultaneously. But it is worth noting that large
contributions from the DOZI processes are also found in the
interpretation~\cite{close-zhao} of the data for $J/\psi\to\omega
f_0(1710)$, $\phi f_0(1710)$, $\omega f_0(1370)$ and $\phi
f_0(1370)$~\cite{bes-phi,bes-plb}. The branching ratio for
$f_0(1710)$ recoiled by $\omega$ in the $J/\psi$ decays is found
to be larger than it being recoiled by $\phi$, while branching
ratio for $\phi f_0(1370)$ is larger than $\omega f_0(1370)$.
Since $f_0(1710)$ is coupled to $K\bar{K}$ strongly and
$f_0(1370)$ prefers to couple to $\pi\pi$ than $K\bar{K}$, a
simple assumption for these two states is that $f_0(1710)$ and
$f_0(1370)$ are dominated by $s\bar{s}$ and $n\bar{n}$,
respectively. Due to this, one would expect that their production
via SOZI processes should be dominant, i.e. $BR(J/\psi\to \phi
f_0(1710))>BR(J/\psi\to \omega f_0(1710))$ and $BR(J/\psi\to
\omega f_0(1370))>BR(J/\psi\to \phi f_0(1370))$. Surprisingly, the
data do not favor such a prescription. In Ref.~\cite{close-zhao},
we find that a glueball-$q\bar{q}$ mixing can explain the scalar
meson decay pattern with a strong contribution from the DOZI
processes. In fact, this should not be out of expectation if
glueball-$q\bar{q}$ mixing occurs in the scalar sector.

We compute two additional decay channels for $\chi_{c0}\to
f_0^if_0^j$, i.e. $\chi_{c0}\to f_0^if_0^j\to
\pi^+\pi^-\pi^+\pi^-$ and $K^+K^-K^+K^-$, which can be examined in
experiment. The results are listed in the last two columns of
Table~\ref{tab-5}. It shows that the largest decay in the $4\pi$
channel is via $f_0(1370)f_0(1500)$, and the smallest channel is
via $f_0(1500)f_0(1710)$. Branching ratios are at order of
$10^{-4}$, the same as the dominant $f_0(1370)f_0(1500)$ channel.
This means that an improved measurement will allow access to most
of those intermediate states if the prescription is correct. In
contrast, decays into four kaons are dominantly via
$f_0(1500)f_0(1710)$ and $f_0(1370)f_0(1710)$ at order of
$10^{-5}$, while all the others are significantly suppressed. The
branching ratio pattern can, in principle, be examined by future
experiment, e.g. at BESIII with much increased statistics.
Nonetheless, uncertainties arising from the $f_0\to PP$ decays can
be reduced.

It should be noted that our treatment for the SU(3) flavor
symmetry breaking in order to reduce the number of free parameters
can be checked by measuring $\chi_{c0}\to a_0(1450)a_0(1450)$. In
the SU(3) symmetry limit, we predict $BR_{\chi_{c0}\to
a_0(1450)a_0(1450)}=5.60\times 10^{-3}$, which is not independent
of $K_0^*(1430)\bar{K_0^*}(1430)$. Experimental information about
this channel will be extremely valuable for clarifying the role
played by the DOZI processes.

In order to examine how this model depends on the scalar mixings,
and learn more about the scalar meson structures, we apply another
two mixing schemes from different approaches and compute the
branching ratios for $\chi_{c0}\to f_0^if_0^j\to \pi^+\pi^-K^+
K^-$, $\pi^+\pi^-\pi^+\pi^-$ and $K^+K^-K^+K^-$. The first one is
from Ref.~\cite{cheng} by Cheng {\it et al.} (Model-CCL) based on
quenched lattice QCD calculations for the glueball spectrum, and
the second one is from Ref.~\cite{giacosa} by Giacosa {\it et al.}
(Model-GGLF) in an effective chiral approach. We note that the
mixing scheme of Ref.~\cite{he-06} with the truncated mixing
matrix for the glueball and $q\bar{q}$ part gives a similar result
as Eq.~(\ref{mix-2}).

In model-CCL, the mix matrix was given as
\bea
\label{ccl} U= \left(
\begin{array}{ccc}
0.859 & 0.302 & 0.413 \\
-0.128 & 0.908 & -0.399 \\
-0.495 & 0.290 & 0.819
\end{array}
\right)  \ .
\eea
With the data from Eqs.~(\ref{kk}) and (\ref{ss13}), we determine
$r=0.90\pm 0.21$. Predictions for other decay channels are given
in Table~\ref{tab-6}.

In Model-GGLF, four mixing solutions were provided. We apply the
first two as an illustration of the effects from the mixing
schemes. The Solution-I gives
\bea
\label{gglf-1} U= \left(
\begin{array}{ccc}
-0.06 & 0.97 & -0.24 \\
0.89 & -0.06 & -0.45 \\
0.45 & 0.24 & 0.86
\end{array}
\right)  \ ,
\eea
and Solution-II reads
\bea
\label{gglf-2} U= \left(
\begin{array}{ccc}
-0.68 & 0.67 & -0.30 \\
0.49 & 0.72 & -0.49 \\
0.54 & 0.19 & 0.81
\end{array}
\right)  \ .
\eea
We then determine $r=1.93\pm 0.29$ and $r=-2.07\pm 0.79$ for
Solution-I and II, respectively. The predictions for the branching
ratios are listed in Tables~\ref{tab-7} and \ref{tab-8}.

Among all these outputs the most predominant feature is that large
DOZI contributions are needed to explain the available data for
$\chi_{c0}\to f_0(1370)f_0(1710)$ and $\chi_{c0}\to
K_0^*(1430)\bar{K_0^*}(1430)$. This also leads to the result that
$\chi_{c0}\to f_0(1370)f_0(1710)\to \pi^+\pi^-K^+K^-$ is a
dominant decay channel. Thinking that all these scalar mixing
schemes have quite different mixing matrix elements, the dominance
of $f_0(1370)f_0(1710)$ gives an impression that the $SS$
branching ratios are not sensitive to the scalar wavefunctions.
However, this is not the case, we note that the data cannot be
explained if $f_0(1710)$ is nearly pure glueball while $f_0(1500)$
a pure $s\bar{s}$, namely, a mixing such as shown by the fourth
solution of Ref.~\cite{giacosa}.

It turns more practical to extract information about the scalar
structures in an overall study of the $SS$ branching ratio pattern
arising from $\chi_{c0}\to SS\to \pi^+\pi^-K^+K^-$, $4\pi$ and
$4K$. For instance, in the $\chi_{c0}\to SS\to 4K$, the dominant
channels are predicted to be via $f_0(1370)f_0(1710)$ and
$f_0(1500)f_0(1710)$ in the mixing of Eq.~(\ref{mix-2}), while in
the other models the $f_0(1500)f_0(1710)$ channel turns out to be
small. In contrast, the $f_0(1370)f_0(1370)$ channel is dominant
in $4\pi$ channel as predicted by Solution-II of Model-GGLF, while
it is compatible with other channels in other solutions.
Systematic analysis of these decay channels should be helpful for
pinning down the glueball-$q\bar{q}$ mixings.

\section{Summary}

A systematic investigation of $\chi_{c0,2}\to VV$, $PP$ and $SS$
in a general parametrization scheme is presented in line with the
new data from BES and CLEO-c. It shows that the exclusive hadronic
decays of the $\chi_{c0,2}$ are rich of information about the
roles played by the OZI-rule violations and SU(3) flavour
breakings in the decay transitions. For $\chi_{c0,2}\to VV$ and
$PP$, we obtain an overall self-contained description of the
experimental data. Contributions from the DOZI processes turn out
to be suppressed. For the channels with better experimental
measurement, i.e. $\chi_{c0,2}\to VV$, and $\chi_{c0}\to PP$, the
SU(3) flavor symmetry is also better respected. Significant SU(3)
breaking turns up in $\chi_{c2}\to PP$ which is likely due to the
poor status of the experimental data and future measurement at
BESIII and CLEO-c will be crucial to disentangle this.

The BES data for $\chi_{c0}\to SS$ allows us to make a
quantitative analysis of the branching ratios in the scalar meson
decay channel. In particular, it allows a test of the scalar $f_0$
mixings motivated by the scalar glueball-$q\bar{q}$ mixing
scenario. Including the new data for $\chi_{c0}\to
K_0^*\bar{K_0^*}$ from BES Collaboration, we find that the decay
of $\chi_{c0}\to f_0^i f_0^j$ favors strong contributions from the
DOZI processes. This phenomenon is consistent with what observed
in $J/\psi\to \phi f_0^i$ and $\omega
f_0^i$~\cite{bes-phi,bes-plb}, where large contributions from the
DOZI processes are also favored~\cite{close-zhao}. The $SS$ decay
branching ratio pattern turns out to be sensitive to the scalar
mixing schemes. An overall study of $\chi_{c0}\to SS\to
\pi^+\pi^-K^+K^-$, $4\pi$ and $4K$ may be useful for us to gain
some insights into the scalar meson structures and extract more
information about the glueball signals in its production channel.

\section*{Acknowledgement}

Useful discussions with C.Z. Yuan and B.S. Zou are acknowledged.
The author is indebted to F.E. Close for many inspiring
discussions. This work is supported, in part, by the U.K. EPSRC
(Grant No. GR/S99433/01), National Natural Science Foundation of
China (Grant No.10675131), and Chinese Academy of Sciences
(KJCX3-SYW-N2).


\begin{table}[ht]
\begin{tabular}{|c|c|c|c|c|}
\hline Parameters &  \multicolumn{2}{c|}{${\chi_{c0}\to VV}$}
&  \multicolumn{2}{c|}{${\chi_{c2}\to VV}$}  \\[1ex]
          \cline{2-5}
& BES & PDG & BES & PDG \\[1ex]\hline
$r$ & $0.203\pm 0.192$ & $0.176\pm 0.197$ & $-0.081\pm 0.098$ & $0.065\pm 0.111$ \\[1ex]
$R$ &  $0.855\pm 0.171$ & $0.825\pm 0.156$ & $0.955\pm 0.148$ & $0.960\pm 0.134$  \\[1ex]
$g_0$ (GeV$^{1/2}$) & $0.291\pm 0.038$ & $0.297\pm 0.042$ &
$0.371\pm 0.039$
& $0.348\pm 0.034$ \\[1ex]
\hline
\end{tabular}
\caption{ The parameters fitted for $\chi_{c0,2}\to VV$ with data
from BES~\cite{bes-99b,bes-04,bes-05} and the world averaged
values from PDG.  } \label{tab-1}
\end{table}

\begin{table}[ht]
\begin{tabular}{|c|c|c|c|c|}
\hline Decay channel &  \multicolumn{2}{c|}{$BR_{\chi_{c0}\to
VV}(\times 10^{-3})$}
&  \multicolumn{2}{c|}{$BR_{\chi_{c2}\to VV}(\times 10^{-3})$}  \\[1ex]
          \cline{2-5}
& BES & PDG & BES & PDG \\[1ex]\hline
$\phi\phi$ & 1.0 $(1.0\pm 0.6)$ & 0.9 $(0.9\pm 0.5)$ & 2.0 $(2.0\pm 0.82)$ & 1.9 $(1.9\pm 0.7)$ \\[1ex]
$\omega\omega$ & 2.29 $(2.29\pm 0.71)$ & 2.3 $(2.3\pm 0.7)$
& 1.77 $(1.77\pm 0.59)$ & 2.0 $(2.0\pm 0.7)$  \\[1ex]
$K^{*0}\bar{K^{*0}}$ & 1.78 $(1.78\pm 0.48)$ & 1.8 $(1.8\pm 0.6)$
& 4.86 $(4.86\pm 1.04)$ & 3.8 $(3.8\pm 0.8)$ \\[1ex]
$\rho\rho$ & 3.457 & 3.755 & 7.532 & 5.816 \\[1ex]
$\omega\phi$ & 0.148 & 0.112 & 0.065 & $\sim 0$\\[1ex] \hline
\end{tabular}
\caption{ The branching ratios obtained for $\chi_{c0,2}\to VV$ by
fitting the data from BES~\cite{bes-99b,bes-04,bes-05} and PDG
average~\cite{pdg2006}. The data are listed in the bracket.}
\label{tab-2}
\end{table}

\begin{table}[ht]
\begin{tabular}{|c|c|c|}
\hline Parameters &  ${\chi_{c0}\to PP}$
&  ${\chi_{c2}\to PP}$  \\[1ex]
          \hline
$r$ & $-0.120\pm 0.044$ & $-0.216\pm 0.102$  \\[1ex]
$R$ &  $1.035\pm 0.067$ & $0.778\pm 0.067$  \\[1ex]
$g_0$ (GeV$^{1/2}$) & $0.366\pm 0.007$ & $0.283\pm 0.008$  \\[1ex]
\hline
\end{tabular}
\caption{ The parameters fitted for $\chi_{c0,2}\to PP$ by
combining the world-average data from PDG~\cite{pdg2006} and the
newly published data from CLEO-c~\cite{cleo-c}.  } \label{tab-3}
\end{table}

\begin{table}[ht]
\begin{tabular}{|c|c|c|c|c|}
\hline Decay channel &  \multicolumn{2}{c|}{$BR_{\chi_{c0}\to
PP}(\times 10^{-3})$}
&  \multicolumn{2}{c|}{$BR_{\chi_{c2}\to PP}(\times 10^{-3})$}  \\[1ex]
          \cline{2-5}
& fit results & data & fit results & data \\[1ex]\hline
$\eta\eta$ & 2.51 &  $(1.9\pm 0.5) \ \ [3.1\pm 0.67]$ & 0.445 & $[< 0.47]$ \\[1ex]
$\eta^\prime\eta^\prime$ & 1.68 &  $[1.7\pm 0.46]$
& 0.076 &  $[< 0.31]$  \\[1ex]
$K^+K^-$ & 5.57 & $(5.4\pm 0.6) $
& 0.924 & $(0.77\pm 0.14)$  \\[1ex]
$K_s^0K_s^0$ & 2.79 & $(2.8\pm 0.7)$ & 0.463 & $(0.67\pm 0.11)$ \\[1ex]
$\pi\pi$ & 7.25 & $(7.2\pm 0.6)$ & 2.123 & $(2.14\pm 0.25)$ \\[1ex]
$\eta\eta^\prime$ & 0.089 & $[<0.50]$ & 0.095 & $[<0.23]$
\\[1ex]\hline
\end{tabular}
\caption{ The branching ratios obtained for $\chi_{c0,2}\to PP$ by
fitting the world-average data from PDG (quoted in the round
bracket)~\cite{pdg2006} together with the new data from CLEO-c
(quoted in the square bracket)~\cite{cleo-c}. } \label{tab-4}
\end{table}

\begin{table}[ht]
\begin{tabular}{|c|c|c|c|c|c|}
\hline Decay channel &  $BR(\chi_{c0}\to SS)(\times 10^{-3})$ &
$B_0 \ (\times 10^{-4})$ & Exp. data  \ $(\times 10^{-4})$ & $B_1
\ (\times 10^{-4})$
& $B_2 \ (\times 10^{-5})$ \\[1ex]
          \hline
$f_0(1370)f_0(1710)$ & 17.80 & 7.12 & $(7.12\pm
1.46\begin{array}{c} +3.28 \\ -1.68\end{array})$ & 1.04 & 5.34 \\[1ex]
$f_0(1370)f_0(1370)$ & 13.14 & 0.17 & $<2.9$ & 2.33 & 0.13 \\[1ex]
$f_0(1370)f_0(1500)$ & 10.76 & 0.62 & $<1.8$ & 3.34 & 0.46 \\[1ex]
$f_0(1500)f_0(1370)$ & 10.76 & 0.25 & $<1.4$ & 3.34 & 0.46 \\[1ex]
$f_0(1500)f_0(1500)$ & 5.02 & 0.50 & $<0.55$ & 2.72 & 0.93 \\[1ex]
$f_0(1500)f_0(1710)$ & 6.18 & 4.31 & $< 0.73$ & 0.63 & 7.98 \\[1ex]\hline
\end{tabular}
\caption{ The branching ratios obtained for $BR_{\chi_{c0}\to
SS}$. $B_0\equiv BR(\chi_{c0}\to SS)\cdot BR(S\to \pi^+\pi^-)\cdot
BR(S\to K^+K^-)$ are branching ratios to be compared with the BES
data~\cite{bes-05b}. $B_1$ and $B_2$ are branching ratios of
$\chi_{c0}\to SS\to \pi^+\pi^-\pi^+\pi^-$ and $\chi_{c0}\to SS\to
K^+K^-K^+K^-$, respectively.} \label{tab-5}
\end{table}

\begin{table}[ht]
\begin{tabular}{|c|c|c|c|c|c|}
\hline Decay channel &  $BR(\chi_{c0}\to SS)(\times 10^{-3})$ &
$B_0 \ (\times 10^{-4})$ & Exp. data  \ $(\times 10^{-4})$ & $B_1
\ (\times 10^{-4})$
& $B_2 \ (\times 10^{-5})$ \\[1ex]
          \hline
$f_0(1370)f_0(1710)$ & 17.80 & 7.12 & $(7.12\pm
1.46\begin{array}{c} +3.28 \\ -1.68\end{array})$ & 1.04 & 5.34 \\[1ex]
$f_0(1370)f_0(1370)$ & 5.06 & 0.07 & $<2.9$ & 0.90 & 0.05 \\[1ex]
$f_0(1370)f_0(1500)$ & 0.04 & $\sim 0$ & $<1.8$ & 0.01 & $\sim 0$ \\[1ex]
$f_0(1500)f_0(1370)$ & 0.04 & $\sim 0$ & $<1.4$ & 0.01 & $\sim 0$ \\[1ex]
$f_0(1500)f_0(1500)$ & 2.43 & 0.24 & $<0.55$ & 1.31 & 0.45 \\[1ex]
$f_0(1500)f_0(1710)$ & 0.74 & 0.52 & $< 0.73$ & 0.08 & 0.96 \\[1ex]\hline
\end{tabular}
\caption{ The branching ratios obtained for $BR_{\chi_{c0}\to SS}$
in Model-CCL~\cite{cheng}. The notations are the same as
Table~\ref{tab-5}. } \label{tab-6}
\end{table}

\begin{table}[ht]
\begin{tabular}{|c|c|c|c|c|c|}
\hline Decay channel &  $BR(\chi_{c0}\to SS)(\times 10^{-3})$ &
$B_0 \ (\times 10^{-4})$ & Exp. data  \ $(\times 10^{-4})$ & $B_1
\ (\times 10^{-4})$
& $B_2 \ (\times 10^{-5})$ \\[1ex]
          \hline
$f_0(1370)f_0(1710)$ & 17.80 & 7.12 & $(7.12\pm
1.46\begin{array}{c} +3.28 \\ -1.68\end{array})$ & 1.04 & 5.34 \\[1ex]
$f_0(1370)f_0(1370)$ & 97.15 & 1.29 & $<2.9$ & 17.27 & 0.97 \\[1ex]
$f_0(1370)f_0(1500)$ & 4.58 & 0.26 & $<1.8$ & 1.42 & 0.20 \\[1ex]
$f_0(1500)f_0(1370)$ & 4.58 & 0.11 & $<1.4$ & 1.42 & 0.20 \\[1ex]
$f_0(1500)f_0(1500)$ & 1.12 & 0.11 & $<0.55$ & 0.61 & 0.21 \\[1ex]
$f_0(1500)f_0(1710)$ & 0.22 & 0.15 & $< 0.73$ & 0.22 & 0.28 \\[1ex]\hline
\end{tabular}
\caption{ The branching ratios obtained for $BR_{\chi_{c0}\to SS}$
with Solution-I of Model-GGLF~\cite{giacosa}. The notations are
the same as Table~\ref{tab-5}. } \label{tab-7}
\end{table}

\begin{table}[ht]
\begin{tabular}{|c|c|c|c|c|c|}
\hline Decay channel &  $BR(\chi_{c0}\to SS)(\times 10^{-3})$ &
$B_0 \ (\times 10^{-4})$ & Exp. data  \ $(\times 10^{-4})$ & $B_1
\ (\times 10^{-4})$
& $B_2 \ (\times 10^{-5})$ \\[1ex]
          \hline
$f_0(1370)f_0(1710)$ & 17.80 & 7.12 & $(7.12\pm
1.46\begin{array}{c} +3.28 \\ -1.68\end{array})$ & 1.04 & 5.34 \\[1ex]
$f_0(1370)f_0(1370)$ & 5.19 & 0.07 & $<2.9$ & 0.92 & 0.05 \\[1ex]
$f_0(1370)f_0(1500)$ & 2.09 & 0.12 & $<1.8$ & 0.65 & 0.09 \\[1ex]
$f_0(1500)f_0(1370)$ & 2.09 & 0.05 & $<1.4$ & 0.65 & 0.09 \\[1ex]
$f_0(1500)f_0(1500)$ & 2.45 & 0.24 & $<0.55$ & 1.33 & 0.45 \\[1ex]
$f_0(1500)f_0(1710)$ & 0.53 & 0.37 & $< 0.73$ & 0.05 & 0.68 \\[1ex]\hline
\end{tabular}
\caption{ The branching ratios obtained for $BR_{\chi_{c0}\to SS}$
with Solution-II of Model-GGLF~\cite{giacosa}. The notations are
the same as Table~\ref{tab-5}. } \label{tab-8}
\end{table}

\end{document}